\documentclass[12pt]{article}
\usepackage{amsmath}
\usepackage{graphicx}

\begin{document}

\title{Renormalisation in Quantum Mechanics}
\author{Michael C. Birse\\
Theoretical Physics Group, School of Physics and Astronomy,\\
The University of Manchester, Manchester, M13 9PL, UK}
\maketitle

\abstract{This lecture provides an introduction to the renormalisation group
as applied to scattering of two nonrelativistic particles. As well as forming 
a framework for constructing effective theories of few-nucleon systems, these
ideas also provide a simple example which illustrates general features of the 
renormalisation group.}

\section{Effective theories}

As particle and nuclear physicists, we are familiar with renormalisation in 
quantum field theory. We meet it first as a trick to get rid of mathematically 
unpleasant divergences. Later we learn to see it as part of a larger structure 
based on scale-dependence: the renormalisation group (RG). This 
is also how it appears in condensed-matter physics, in the context of critical 
phenomena \cite{wilson}.

The same ideas can also be used to study scale dependence in much simpler 
systems: just two or three nonrelativistic particles. They are of particular 
interest in nuclear physics, where we are trying to construct systematic 
effective field theories of nuclear forces (see \cite{reviews} for recent
reviews). They can also be applied to systems of cold atoms in traps, where 
magnetic fields can be used to tune the interactions between the atoms.
In addition, these applications provide tractable examples of RG flows. 
Without the complications of a full field theory, the equations can often 
be solved exactly while still illustrating all of the general features of
these flows \cite{bmr}.

Effective field theories describe only the low-energy degrees of freedom of some 
system and so they are not ``fundamental". In general they are not renormalisable
and so they contain an infinite number of terms. This is potentially a disaster 
for their predictive power, but not if we can find a systematic way to organise 
these terms. Then, at any order in some expansion, only a finite number of terms 
will contribute. Having determined the coefficients of these by fitting them to 
data (or to simulations of the underlying physics), we can use them to predict 
other observables.

\begin{figure}[h]
\centering
\includegraphics[width=6cm]{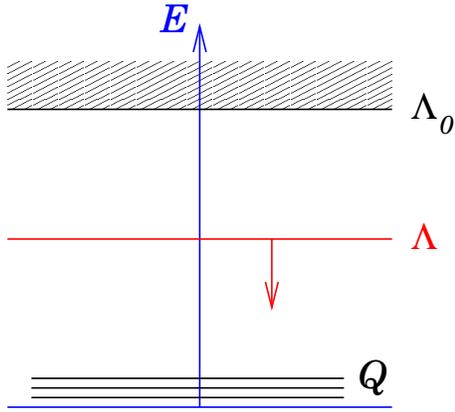}
\caption{Scales and the running cut-off.}
\end{figure}

This works provided there is a good separation of scales, as illustrated in Fig.~1. 
Here $Q$ generically denotes the experimentally relevant low-energy scales and
$\Lambda_0$ the scales of the underlying physics. In the case of nuclear physics,
the low scales include particles' momenta and the pion mass, while the high scales 
include the scale of chiral symmetry breaking, $4\pi f_\pi$, and the masses of 
hadrons like the $\rho$ meson and nucleon. If these are well separated, we can expand 
observables in powers of the small parameter $Q/\Lambda_0$. The terms in the 
effective theory can then be organised according to a ``power counting" in the low 
scales $Q$.

The effective theory describes physics at low momenta. Short-range physics is not 
resolved by it and so is just represented by contact interactions 
($\delta$-functions and their derivatives). However scattering by these is 
ill-defined since they couple to virtual states with arbitrarily high momenta. 
The basic nonrelativistic loop diagram (which is relevant for the rest of this 
talk) is shown in Fig. 2. For $S$-wave scattering this integral is
\begin{equation}
M\int \frac{q^2\,{\rm d}q}{p^2-q^2}\sim -M\int {\rm d}q,\quad\mbox{for large\ }q,
\end{equation} 
and so contains a linear divergence. We therefore need to regulate the theory. There 
are many ways to do this: dimensional regularisation \cite{ksw}, a simple 
momentum cut-off \cite{bmr}, or adding a term to the kinetic energy to suppress 
high-energy modes \cite{hkn}. All of these are equivalent, but each introduces some 
arbitrary scale, $\Lambda$. This is essentially the highest momentum that is 
included explicitly in the theory. Physical predictions should be independent of 
$\Lambda$ and this leads us to the RG.

\begin{figure}[h]
\centering
\includegraphics[width=3cm]{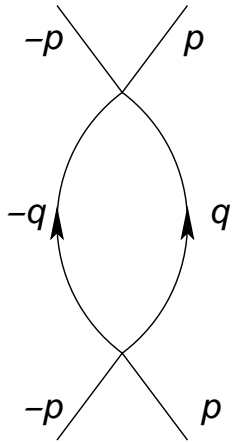}
\caption{The basic loop integral.}
\end{figure}

As we lower $\Lambda$, the couplings must run. This is because more and more 
physics is ``integrated out" (see Fig.~1) and so must be included implicitly in 
the effective couplings. Ultimately we lose all memory of the underlying physics 
and the only scale we have left is $\Lambda$. In units of $\Lambda$, everything 
is then just a number. We have arrived at a fixed point of the RG -- a scale-free 
system. These are the end-points of the RG flow. Two are shown in Fig.~2. The one 
on the left is stable: any nearby theory will flow towards it as the the cut-off 
is lowered. In contrast, the one on the right has an unstable direction: the flow 
can take theories away from the fixed point unless they lie on the ``critical 
surface".

\begin{figure}[h]
\centering
\includegraphics[width=8cm]{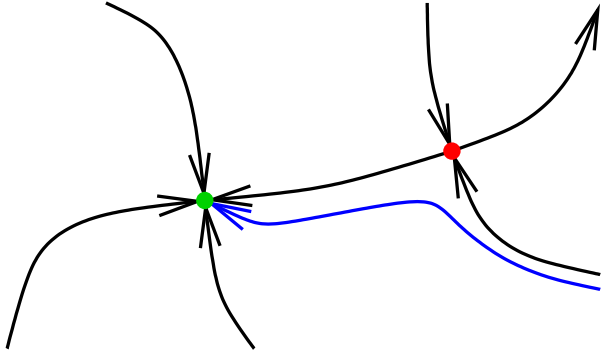}
\caption{Fixed points.}
\end{figure}

Close to a fixed point, we can find perturbations that show a power-law dependence
on $\Lambda$ and we can use this power counting  to organise
the terms in our effective theory. They can be classified into three types:
\begin{itemize}
\item $\Lambda^{-\nu}$: relevant/super-renormalisable\footnote{The term
``relevant" is commonly used in condensed-matter physics, whereas 
``super-renormalisable" is more usual in particle physics.},\newline
for example mass terms in quantum field theories like QED;
\item $\Lambda^0$: marginal/renormalisable,\newline
for example the couplings familiar in gauge theories like the Standard Model
(typically these show a $\log\Lambda$ dependence on the cut-off);
\item $\Lambda^{+\nu}$: irrelevant/nonrenormalisable,\newline
for example the interactions in Chiral Perturbation Theory.
\end{itemize}

\section{RG equation for two-body scattering}

Let us look at scattering of two non-relativistic particles at low enough
energies that the range of the forces is not resolved (for example, two nucleons
with an energy below about 10~MeV). This can be described by an effective 
Lagrangian with two-body contact interactions or, equivalently, a Hamiltonian
with a $\delta$-function potential. In momentum space, the $S$-wave potential can 
be written
\begin{equation}
V(k',k,p)=C_{00}+C_{20}(k^2+k'^2)+C_{02}\,p^2\cdots,
\end{equation}
where $k$ and $k'$ denote the initial and final relative momenta and the 
energy-dependence is expressed in terms of the on-shell
momentum $p=\sqrt{ME}$. 

Scattering can be described by the reactance matrix ($K$), defined similarly to 
the scattering matrix ($T$) but with standing-wave boundary conditions. This has 
the advantage that it is real below the particle-production threshold. For 
$S$-wave scattering, it satisfies the Lippmann-Schwinger equation
\begin{equation} 
K(k',k,p)=V(k',k,p)
+\frac{M}{2\pi^2}\,{\cal{P}}\!
\int_{0}^{\Lambda}q^2{\rm d}q\,\frac{V(k',q,p)K(q,k,p)}
{{p^2}-{q^2}},
\end{equation}
where ${\cal P}$ denotes the principal value. This integral equation sums 
chains of the bubble diagrams in Fig.~2 to all orders. On-shell 
($k'=k=p$), the $K$-matrix is related to the $T$-matrix by
\begin{equation}
\frac{1}{K(p)}=\frac{1}{T(p)}-{\rm i}\,\frac{Mp}{4\pi}
=-\,\frac{Mp}{4\pi}\,\cot\delta(p),
\end{equation}
where $\delta(p)$ is the phase shift.

With contact interactions, the integral over the momentum $q$ of the 
virtual states is divergent and so we need to regulate it. Here I
follow the method developed in \cite{bmr} and simply cut the
integral off at $q=\Lambda$. We can write the integral equation in the 
schematic form
\begin{equation}
K=V+VGK.
\end{equation}
Demanding that the off-shell $K$-matrix be independent of $\Lambda$,
\begin{equation}
\dot K\equiv \frac{\partial K}{\partial\Lambda}=0,
\end{equation}
ensures that scattering observables will be independent of the
arbitrary cut-off. Differentiating the integral equation gives
\begin{equation}
0=\dot V+\dot VGK+V\dot GK,
\end{equation}
where $\dot G$ implies differentiation with respect to the cut-off on the
integral. Multiplying this by $(1+GK)^{-1}$ and using 
the integral equation for $K$, we arrrive at
\begin{equation}
\dot V=-V\dot GV.
\end{equation}

This equation has a very natural structure: as states at the cut-off, with 
$q=\Lambda$, are removed from the loop integral in Fig.~2, their effects 
are added into the potential to compensate. Written out explicitly, it is
\begin{equation}
\frac{\partial V}{\partial\Lambda} 
=\frac{M}{2\pi^2}\,V(k',\Lambda,p,\Lambda)\,\frac{\Lambda^2}{\Lambda^2-p^2}\,
V(\Lambda,k,p,\Lambda).
\end{equation}
Note that the use of the fully off-shell $K$-matrix was essential to
obtaining an equation involving only the potential; a similar approach
based on the half-off-shell $T$-matrix yields an equation that still 
involves the scattering matrix \cite{bskb}.

This equation for the cut-off dependence of the effective potential is 
still not quite the RG equation: the final step is to express all dimensioned
quantities in units of $\Lambda$. Rescaled momentum variables (denoted with hats)
are defined by $\hat k=k/\Lambda$ etc., and a rescaled potential by
\begin{equation}
\hat V(\hat k',\hat k,\hat p,\Lambda)=\frac{M\Lambda}{2\pi^2}\,
V(\Lambda\hat k',\Lambda\hat k,\Lambda\hat p,\Lambda).
\end{equation}
(The factor $M$ in this corresponds to dividing an overall factor of $1/M$ out 
of the Schr\"odinger equation.) This satisfies the RG equation\pagebreak
\begin{eqnarray}
\Lambda\,\frac{\partial\hat V}{\partial\Lambda}&=&
\hat k'\,\frac{\partial\hat V}{\partial\hat k'}
+\hat k\,\frac{\partial\hat V}{\partial\hat k}
+\hat p\,\frac{\partial\hat V}{\partial\hat p}
+\hat V\cr
&&\qquad+\hat V(\hat k',1,\hat p,\Lambda)\,\frac{1}{1-\hat p^2}\,
\hat V(1,\hat k,\hat p,\Lambda).
\end{eqnarray}
The sum of logarithmic derivatives is similar to the structure of analogous
RG equations in condensed-matter physics; it counts the powers of
low-energy scales present in the potential. The boundary conditions on 
solutions to this equation are that they should be analytic functions of 
$\hat k^2$, $\hat k^{\prime 2}$ and $\hat p^2$ (since they should arise from
an effective Lagrangian constructed out of $\partial/\partial t$ and 
$\nabla^2$). For small values of these quantities the potential should thus have
an expansion in non-negative integer powers of them.

\section{Fixed points and perturbations}

Having constructed the RG equation, the first thing we should do is to
look for fixed points -- solutions that are independent of $\Lambda$.
There is one obvious one: the trivial fixed point
\begin{equation}
\hat V=0.
\end{equation}
(Since there is no scattering, this obviously describes a scale-free system.)

To describe more interesting physics, we need to expand around the fixed point,
looking for perturbations that scale with definite powers of $\Lambda$. These
are eigenfunctions of the linearised RG equation. They have the form
\begin{equation}
\hat V(\hat k',\hat k,\hat p,\Lambda)=\Lambda^\nu \phi(\hat k',\hat
k,\hat p),
\end{equation}
and they satisfy the eigenvalue equation
\begin{equation}
\hat k'\,\frac{\partial\phi}{\partial\hat k'}
+\hat k\,\frac{\partial\phi}{\partial\hat k}
+\hat p\,\frac{\partial\phi}{\partial\hat p}
+\phi=\nu\phi.
\end{equation}
Its solutions are
\begin{equation}
\phi(\hat k',\hat k,\hat p)=C\,\hat k^{\prime 2l}\,\hat k^{2m}\,\hat p^{2n},
\end{equation}
with $k,l,m\geq 0$ since only non-negative, even powers satisfy the boundary
condition. The corresponding eigenvalues are
\begin{equation}
\nu=2(l+m+n)+1.
\end{equation}
These are all positive and so the fixed point is stable. The eigenvalues
simply count the powers of low-energy scales. ($\nu=d+1$ where $d$ is
the ``engineering dimension", as in Weinberg's original power counting
for ChPT \cite{wein}.)

There are also many nontrivial fixed points, all of which are unstable. The most
interesting one is purely energy-dependent. To study it, I focus on potentials
of the form $V(p,\Lambda)$. The RG equation for these simplifies to
\begin{equation}
\Lambda\,\frac{\partial\hat V}{\partial\Lambda}
=\hat p\,\frac{\partial\hat V}{\partial\hat p}
+\hat V+\frac{\hat V(\hat p,\Lambda)^2}{1-\hat p^2}.
\end{equation}
Since all terms involve just one function, we can divide by $\hat V^2$
to get
\begin{equation}
\Lambda\,\frac{\partial}{\partial\Lambda}\left(\frac{1}{\hat V}\right)
=\hat p\,\frac{\partial}{\partial\hat p}\left(\frac{1}{\hat V}\right)
-\frac{1}{\hat V}-\frac{1}{1-\hat p^2},
\end{equation}
which is just a linear equation for $1/\hat V(\hat p,\Lambda)$.

To find the fixed point, we set the LHS of this equation to zero. The
resulting ODE can then be integrated easily. The general solution is
\begin{equation}
\frac{1}{\hat V_0(\hat p)}=-\int_0^1\frac{\hat q^2\,{\rm d}\hat q}{\hat q^2-\hat p^2}
+C\hat p.
\end{equation}
The final term is not analytic in $\hat p^2$ and so the boundary condition
requires $C=0$. The fixed-point potential is thus
\begin{equation}
\frac{1}{\hat V_0(\hat p)}=-1+\frac{\hat p}{2}\ln\frac{1+\hat p}{1-\hat p}.
\end{equation}
The precise form of this is regulator-dependent (for example, it 
is just a constant for dimensional regularisation \cite{ksw}), but the presence
of a negative constant of order unity is generic.

Since this potential has no momentum dependence, the integral equation for 
the $K$-matrix simplifies to an algebraic equation. In rescaled, dimensionless
form, it can be written
\begin{equation}
\frac{1}{\hat K(\hat p)}=\frac{1}{\hat V_0(\hat p)}-\int_0^1
\frac{\hat q^2\,{\rm d}\hat q}{\hat p^2-\hat q^2}.
\end{equation}
The integral here is just the negative of the one above in $1/\hat V_0$ itself 
and so we get
\begin{equation}
\frac{1}{\hat K(\hat p)}=0.
\end{equation}
The corresponding $T$-matrix,
\begin{equation}
\frac{1}{\hat T(\hat p)}=\frac{1}{\hat K(\hat p)}+{\rm i}\,\frac{\pi}{2}\,\hat p,
\end{equation}
has a pole at $\hat p=0$. The fixed-point therefore describes a system with a 
bound state at exactly zero energy (another scale-free system). 

More general systems can be described by perturbing around the fixed point. 
In particular, energy-dependent perturbations can be found by substituting 
\begin{equation}
\frac{1}{\hat V(\hat p,\Lambda)}=\frac{1}{\hat V_0(\hat p)}
+\Lambda^\nu\phi(\hat p)
\end{equation}
into the RG equation. The functions $\phi(\hat p)$ satisfy the eigenvalue 
equation
\begin{equation}
\hat p\,\frac{\partial\phi}{\partial\hat p}-\phi=\nu\phi.
\end{equation}
The solutions to this are powers of the energy,
\begin{equation}
\phi(\hat p)=C\hat p^{2n},
\end{equation}
with eigenvalues
\begin{equation}
\nu=2n-1.
\end{equation}
The RG eigenvalues for these perturbations have been shifted by $-2$ compared
to the simple ``engineering" power counting. There is one negative eigenvalue
and so the fixed point is unstable. 

A slice through the RG flow is shown in Fig.~4. The two fixed points can be seen, 
as well as the critical line through the nontrivial one. Potentials close to this 
line initially flow towards the fixed point as we lower the cut-off but are then 
diverted away from it. A potential to the right of the line is not quite strong 
enough to produce a bound state. As $\Lambda$ passes through the scale associated 
with the virtualstate, the flow turns to approach the trivial fixed point from the 
weakly attractive side. In contrast, a potential to the left of the critical line
generates a finite-energy bound state. This state drops out of our low-energy 
effective theory when the cut-off reaches the corresponding momentum scale. As 
this happens, the RG flow takes the potential to infinity and it then reappears 
from the right, ultimately approaching the trivial fixed point from the weakly 
repulsive side.

\begin{figure}[h]
\centering
\includegraphics[width=11cm]{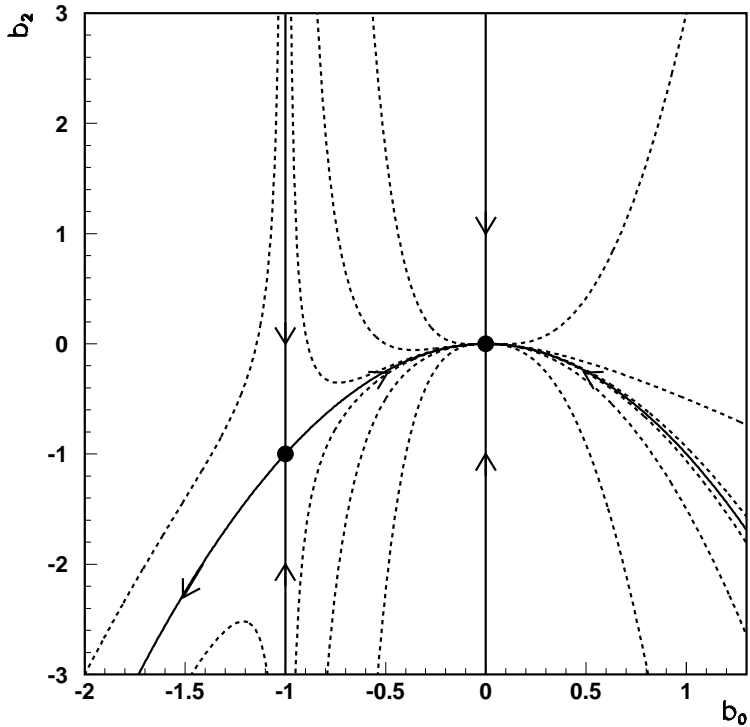}
\caption{RG flow of the potential $\hat V(\hat p,\Lambda)
=b_0(\Lambda)+b_2(\Lambda)\,\hat p^2+\cdots$.}
\end{figure}

\noindent \textbf{Exercise:} \textit{Repeat this analysis for a general 
number of space dimensions, in particular for $D=1$ and 2, and interpret your 
results.}
\vspace{5pt}

Physical observables are given by the on-shell $K$-matrix. Returning to
physical units, this is
\begin{equation}
\frac{1}{K(p)}=\frac{M}{2\pi^2}\sum_{n=0}^\infty C_n\, p^{2n},
\end{equation}
where the $C_n$ are the coefficients of the RG eigenfunctions in $1/\hat V$.
Comparing this with
\begin{equation}
\frac{1}{K(p)}=-\,\frac{Mp}{4\pi}\left(-\,\frac{1}{a}
+\frac{1}{2}\,r_ep^2+\cdots\right),
\end{equation}
we see that this expansion is, in fact, just the effective-range expansion 
(first applied to the nucleon-nucleon interaction by Bethe in 1949 \cite{bethe}).
Note that the terms in the expansion of our effective theory have a direct
connection to scattering observables. This is as it should be: effective theories
are systematic tools to analyse data, not fundamental theories that aim to 
predict everything in terms of a small number of parameters.

Finally, I should make a brief comment about momentum-dependent perturbations
around the nontrivial fixed point, which I have not discussed above. These terms 
change the off-shell dependence of the scattering matrix, without affecting
physical observables. Their explicit forms can be found in Ref.~\cite{bmr}. In 
contrast to the expansion around the trivial fixed point, momentum- and 
energy-dependent terms appear at different orders. Specifically, the 
momentum-dependent perturbations around the nontrivial point have even RG 
eigenvalues. Each term is one order higher in the expansion than the 
corresponding energy-dependent one. This means that using them to eliminate
energy dependence will leave an effective potential without an obvious power 
counting (like the potential obtained in Ref.~\cite{bskb}).

\section{Extensions}

Here I have discussed only the application of the RG to systems where
the range of the forces is not resolved and the interactions 
can all be represented by contact terms. There are many other systems with known
long-range forces, for example: Coulomb, pion exchange, dipole-dipole or van der 
Waals interactions. Similar RG methods can be applied to the unresolved 
short-range forces accompanying these \cite{bb1,birse}. The resulting 
expressions are either distorted-wave Born expansions or distorted-wave versions 
of the effective-range expansion. (In the case of the Coulomb potential, it was 
again Bethe who first wrote this expansion down \cite{bethe}.)

Another important application is to the $1/r^2$ potential that arises in
three-body systems with attractive short-range forces \cite{bb2}. If the 
two-body scattering length is infinite, the Efimov effect leads to a tower of 
geometrically-spaced bound states \cite{efimov}. This is the origin of the 
limit cycle that has been found in the RG flows for these systems \cite{bhvk}
(one of the few known examples of such a cycle).

\section*{Acknowledgments}

I am grateful to M. Rosina, B. Golli and S. Sirca for the invitation to
participate in the Bled 2007 Workshop ``Hadron structure and lattice QCD".
I should also thank them and L. Glozman for providing the impetus to write 
up this lecture. Finally, I acknowledge the contributions of my
collaborators, J. McGovern, K. Richardson and T. Barford, to the work
outlined here.


\begin{thebibliography}{99}
\bibitem{wilson}K. G. Wilson, Rev. Mod. Phys. \textbf{55} (1983) 583.
\bibitem{reviews}P. F. Bedaque and U. van Kolck, Ann. Rev. Nucl. Part. Sci.
\textbf{52} (2002) 339 [nucl-th/0203055];
E. Epelbaum, Prog. Part. Nucl. Phys. \textbf{57} (2006) 654 [nucl-th/0509032].
\bibitem{bmr}M. C. Birse, J. A. McGovern and K. G. Richardson, 
Phys. Lett. \textbf{B464}, 169 (1999) [hep-ph/9807302].
\bibitem{ksw}Nucl. Phys. \textbf{B534} (1998) 329 [nucl-th/9802075].
\bibitem{hkn}K. Harada, H. Kubo and A. Ninomiya, nucl-th/0702074.
\bibitem{bskb}S. K. Bogner, A. Schwenk, T. T. S. Kuo and G. E. Brown,
nucl-th/0111042; see also: S. K. Bogner \textit{et al.}, Phys. Lett. 
\textbf{B576} (2003) 265 [nucl-th/0108041].
\bibitem{wein}S. Weinberg, Physica \textbf{A96} (1979) 327;
Phys. Lett. {\bf B251} (1990) 288.
\bibitem{bethe}H. A. Bethe, Phys. Rev. \textbf{76} (1949) 38.
\bibitem{bb1}T. Barford and M. C. Birse, Phys. Rev. \textbf{C67} (2003) 064006
[hep-ph/0206146].
\bibitem{birse}M. C. Birse, Phys. Rev. \textbf{C74} (2006) 014003 
[nucl-th/0507077].
\bibitem{bb2}T. Barford and M. C. Birse, J. Phys. A: Math. Gen. \textbf{38} 
(2005) 697 [nucl-th/0406008].
\bibitem{efimov}V. N. Efimov, Sov. J. Nucl. Phys. \textbf{12} (1971) 589;
\textbf{29} (1979) 546.
\bibitem{bhvk}P. F. Bedaque, H.-W. Hammer and U. van Kolck,
Phys. Rev. Lett. \textbf{82}, 463 (1999) [nucl-th/9809025];
Nucl. Phys. \textbf{A646}, 444 (1999) [nucl-th/9811046];
Nucl. Phys. \textbf{A676}, 357 (2000) [nucl-th/9906032].


\end{thebibliography}
\end{document}